\documentstyle[proceedings]{crckapb} 

\begin{opening}
\title{VLBA MONITORING OF THE QSO 3C\,345}
\author{E. ROS}
\author{J.A. ZENSUS}
\author{A.P. LOBANOV}
\institute{Max-Planck-Institut f\"ur Radioastronomie, 
           Bonn, Germany}
\end{opening}

\runningtitle{VLBA MONITORING OF THE QSO 3C\,345}

\begin{document}


\section{Introduction.}
The QSO 3C\,345 ($V$=16, $z$=0.595) is a core-dominated radio
source that displays apparent superluminal motions,
with components traveling in the parsec-scale jet along
curved trajectories and speeds up to 10$c$ \cite{zen95}.
We have observed this QSO with the very
long baseline interferometry (VLBI) technique using the NRAO Very Long
Baseline Array (VLBA) at three epochs and four frequencies, in
order to monitor it in total and linear polarization intensity.

\begin{figure}[htb]
\vspace{140mm}
\includegraphics{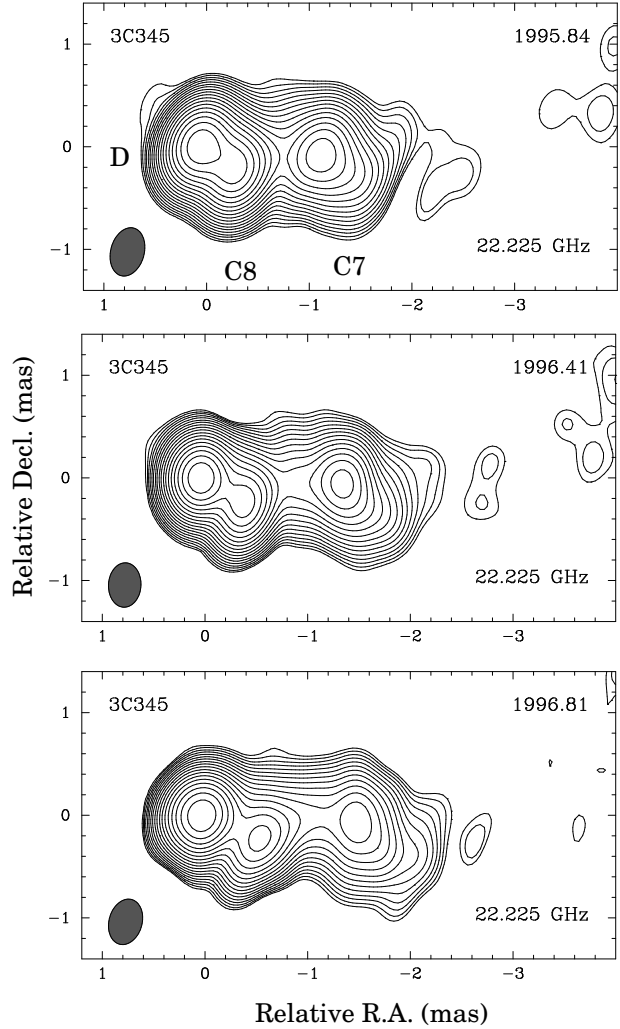}
\caption{Total intensity images of the QSO 3C\,345 at 22\,GHz for the
epochs 1995.84, 1996.41, and 1996.81.  The
synthesized beams are, respectively, 
0.480$\times$0.326, P.A.\,-15$\rlap{.}^\circ$0, 
0.433$\times$0.316, P.A.\,-3$\rlap{.}^\circ$1, and
0.451$\times$0.322, P.A.\,-15$\rlap{.}^\circ$5.  The total fluxes recovered
in the maps are 6.98, 4.51 and 4.61\,Jy.  The peaks
of brightness are 2.56, 1.80, and 2.09\,Jy/beam.  The contours displayed
are, 5.1, 3.7, and 4.2\,mJy\,$\times$(-1,1.4,...,256,362).}
\label{fig:maps22}
\end{figure}

\section{Observations and Imaging.}

We observed 3C\,345 in 1995.84, 1996.41, and 1996.81, using the VLBA at 22, 15,
8.4, and 5\,GHz, and recording with a 16\,MHz bandwidth at all frequencies.
At each frequency, the source was observed for about 14\,hrs, using 5-minute
scans and interleaving all observing frequencies.  Some calibrator scans 
(on 3C\,279, 3C\,84, NRAO\,91, OQ\,208 and 3C\,286)
were inserted during the observations.

\paragraph{Total Intensity.}
After the fringe-fitting process, we exported the data into the differential
mapping program DIFMAP \cite{she94}, and we obtained total intensity
images using the hybrid mapping technique.
The components C7 and C8 can be identified 
close to the core D, in the images at higher frequencies.  
At the lower frequencies, the jet extends to the NW direction, turning to the
N at $\ge$20\,mas distances from the core.
In Fig.\ \ref{fig:maps22}, we
show the central region of the source at 22\,GHz for the three
sampled epochs.

To describe the structures observed within $\sim$3\,mas distance from
the core, we fit
elliptical components
with Gaussian brightness profiles to the visibility data at 15 and 22 GHz.
The change measured in the fitted positions for the component C8 indicates
that the component was moving away from the core, at an angular speed
of 0.26$\pm$0.08\,mas/yr at 22\,GHz, which corresponds to an apparent speed
of 5.1$\pm$1.8$h^{-1}\,c$
(assuming 
$H_0$=100$h$\,km\,s$^{-1}$ Mpc$^{-1}$ and $q_0$=0.5). 
The trajectory of C7 at the same frequency shows a proper motion of
0.29$\pm$0.08\,mas/yr, or apparent speed of 5.7$\pm$1.8$h^{-1}\,c$.
At 15\,GHz the respective values are 
C8:\,$0.23\pm0.11$\,mas/yr ($4.5\pm2.2\,h^{-1}c$) 
and C7:\,$0.30\pm0.11$\,mas/yr ($5.9\pm2.2\,h^{-1}c$).

\begin{figure}[htb]
\vspace{70mm}
\includegraphics{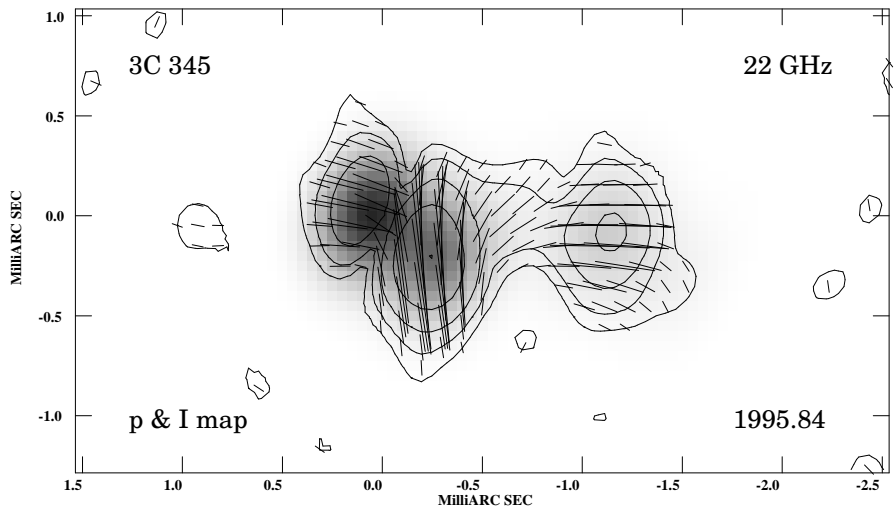}
\caption{VLBA $p$, $I$ and $\chi$ images of 3C\,345 at 22\,GHz, 
epoch 1995.84.  The polarized intensity $p$ is represented with contours 
(value of 7\,mJy/beam$\times$(1,2,4,8,16), brightness peak of 112\,mJy/beam)
over a grey scale total intensity map (grey scale above 7\,mJy/beam,
check Fig.\ 1, top for parameters), and the superimposed electric vectors 
($\chi$, length $\propto$$p$).}
\label{fig:pol22}
\end{figure}

\paragraph{Polarized Intensity.}
Since the early works of \citeauthor{cot84} \shortcite{cot84} up to now, 
important progress has been made in polarimetric VLBI observations.
The VLBA has standardized feeds with low
instrumental polarization, and the relatively small size of the antennas
is compensated by the excellent performance of the receivers.  In addition,
a new method for self-calibrating the polarimetric data has been introduced
\cite{lep95}, enabling D-term determination by using the program source 
itself.
Using this method, we have calibrated
the instrumental polarization, determining the feed solutions for the receivers
of all the antennas.  
This has allowed us to obtain maps of the linearly polarized emission from
the source
($P=Q+iU=pe^{2i\chi}=
mIe^{2i\chi}$, where $Q$ and $U$ are the Stokes parameters,
$p=mI$ is the polarized intensity, $m$ is the fractional
linear polarization, and $\chi$ is the position angle of
the electric vector in the sky). 
In Fig.\ \ref{fig:pol22}, we show a composition of the total intensity 
$I$ (grey scale), the polarized intensity $p$ (contours) and the electric 
vector orientation angle $\chi$ (segments, length
proportional to p) for the 22\,GHz observations in 1995.84.

In Fig.\ \ref{fig:pol22} the electric vector is 
aligned with the extremely curved jet direction at the inner 3\,mas both
for the core and the C7 component.  The component C8 is boosted in polarized 
emission, showing an electric vector apparently perpendicular to this jet 
direction.
\citeauthor{gom94a} \shortcite{gom94a,gom94b} have modeled features
similar to the ones presented here.  They reported
anticorrelation between the polarized and the total intensity flux
close to a shock wave along a jet in its early evolution (near to the
core).
We should point out that the possible superposition between components with 
different electric vector orientations can lead to a cancellation
of flux and produce apparent 
separations between polarized emission components.  The brighter polarized 
emission in C8 might be explained by a shock wave in a curved jet geometry.
More generally, the features of Fig.\ \ref{fig:pol22} can be explained in
terms of a comprehensive shock model \cite{war94} in the framework 
of a helical geometry for the motion of the components \cite{ste95}.

%

\section{Conclusions.}

We have monitored the superluminal QSO 3C\,345 at three epochs within one year,
observing with the VLBA at four frequencies.
We have presented some 
results of these studies at the higher frequencies, showing the 
superluminal motions of components C8 and C7 with respect to 
the core component D, and the remarkably complex polarization structure
near the core, which provides evidence
for emerging components and changing projected jet direction 
within 3\,mas from the core.
The twist in the orientation of the electric vector along
the jet can be explained in terms of
an extremely curved helical geometry, following \citeauthor{ste95}
\shortcite{ste95}. 
The electric field is parallel to the jet direction, and the 
boosting in the polarized emission at the component C8 and the change
in the vector orientation can be the result of the presence of a shock
wave in a bent jet.




\end{document}